# MUON LOCALIZATION SITE IN U(Pt,Pd)$_3$

A. Schenck[a], R.J. Keizer[b], A. de Visser[b], A. Amato[c], D. Andreica[a], F.N. Gygax[a], M. Pinkpank[a], P. Estrela[b], M.J. Graf[d], A.A. Menovsky[b] and J.J.M. Franse[b]

[a]*Institute for Particle Physics, ETH Zürich, PSI, CH-5232 Villigen, Switzerland*

[b]*Van der Waals-Zeeman Institute, University of Amsterdam, Valckenierstraat 65, 1018 XE Amsterdam, The Netherlands*

[c]*Paul Scherrer Institute, CH-5232 Villigen, Switzerland*

[d]*Department of Physics, Boston College, Chestnut Hill, MA 02467, USA*

**Abstract**

The angular and temperature (10-250 K) variation of the Knight shift of single-crystalline U(Pt$_{0.95}$Pd$_{0.05}$)$_3$ has been measured in transverse field ($B$=0.6 T) μSR experiments. By analysing the temperature variation of the Knight shift with a modified Curie-Weiss expression the muon localization site in this hexagonal material is determined at (0,0,0).

Keywords: muon localization site, heavy-fermions, U(Pt$_{0.95}$Pd$_{0.05}$)$_3$

Address correspondence to:
    Dr. A. de Visser
    Van der Waals-Zeeman Institute, University of Amsterdam
    Valckenierstraat 65, 1018 XE Amsterdam, The Netherlands
    Phone: +31-20-5255732, Fax: +31-20-5255788
    E-mail: devisser@wins.uva.nl

## 1. Introduction

The heavy-fermion material UPt$_3$ presents an exemplary system to study the antiferromagnetic instability in strongly correlated electron systems. Pure UPt$_3$ orders antiferromagnetically at a Néel temperature $T_N \sim 6$ K, with an extremely small ordered moment ($m= 0.02$ $\mu_B$/U-atom). A recent neutron-diffraction study [1] shows that this unusual type of small-moment magnetism is robust when Pt is replaced by isoelectronic Pd. The small ordered moment in the U(Pt$_{1-x}$Pd$_x$)$_3$ pseudobinaries grows with increasing $x$ ($x \leq 0.01$), while $T_N$ remains ~ 6 K. In between $x= 0.005$ and $x= 0.01$ a second antiferromagnetic phase emerges, which is no longer observed for $x \geq 0.10$. This phase presents a more conventional type of magnetism with a maximum $T_N$ of 5.8 K and a substantial ordered moment of 0.6 $\mu_B$/U-atom for $x= 0.05$. The magnetic structures of the small-moment and large-moment antiferromagnetic phases, as determined by neutron diffraction, are identical. In order to study the evolution of magnetism in the U(Pt,Pd)$_3$ pseudobinaries we have carried out a series of zero- and transverse field µSR experiments [2-5]. Determination of the muon localization site in this hexagonal material is the topic of this paper.

## 2. Experimental

µSR experiments were carried out on single-crystalline U(Pt$_{0.95}$Pd$_{0.05}$)$_3$ in the General Purpose Spectrometer at the PSI. Transverse field ($B= 0.6$ T) data were taken on two single crystals prepared by the Czochralski technique. One sample (#1) was shaped into a sphere (diameter 3 mm) and a second sample (#2) into a cube with edges (length 5 mm) along the principal crystallographic axes (space group P6$_3$/mmc), denoted by a, a* and c, where a*⊥a in the basal plane. The Knight shift, $K_a(T)$ and $K_c(T)$, was measured in the temperature range 10-250 K, while the angular dependence $K(\theta)$ was determined for fields directed in the a-c and a-a* plane at $T=10$ K. Sample #2 was mounted on a silver support which gave rise to an additional background signal. Sample #1 was mounted with mylar tape on a copper fork-like holder, which resulted in a significant reduction of the background signal.

## 3. Results

One of the most striking results of the TF experiments on U(Pt$_{0.95}$Pd$_{0.05}$)$_3$ is that for **B**||**a** two closely spaced frequency components are observed, but for **B**||**c** only one. At the lowest temperature, $T$= 10 K, the two components clearly show up in the Fourier transform (albeit for sample #2 only) with equal weight [5]. For $T$> 10 K, the two signals in the Fourier transform for **B**||**a** are no longer resolved and instead a single asymmetric peak is observed. For **B**||**c** we have analysed the spectra with a standard two-component depolarisation function (due to the sample and background), while for **B**||**a** we have analysed the spectra using the following three-component depolarisation function, $G_a(t)$:

$$G_a(t) = A_1 e^{-\lambda_1 t} \cos(2\pi\nu_1 t + \varphi) + A_2 e^{-\lambda_2 t} \cos(2\pi\nu_2 t + \varphi) \\ + A_{bg} e^{-\lambda_{bg} t} \cos(2\pi\nu_{bg} t + \varphi) \quad (1)$$

Here the two frequency ($\nu_1$, $\nu_2$) components are exponentially damped and the last term accounts for the background signal. Although, for $T$> 10 K, the first two signals are not resolved in the frequency domain, it is possible to fit both components in the time domain by fixing the asymmetries $A_1 = A_2$. In Fig.1 $K_a(T)$ and $K_c(T)$ are plotted. For **B**||**a** we show $K_a(T)$ obtained by fitting a one or a two-component function in addition to the background signal. For $T$> 100 K the difference between the two frequencies reduces and above 150 K the data can only be fitted with one frequency. This indicates that slow muon hopping takes place for $T$> 100 K. When the muon diffuses, it experiences an average local magnetic field.

In Fig.2, we show the angular variation of the Knight shift frequency measured on the spherical sample in the a-a* and a-c plane at $T$= 10 K. For simplicity we show for **B**||**a** the results of the analysis with one frequency component only. In the a-a* plane $K(\theta)$ is isotropic, while $K(\theta)$ follows the standard cos$^2$-law for $B$ directed in the a-c plane. From these symmetry constraints we conclude that the muon localization site is restricted to axial symmetry.

## 3. Analysis and discussion

In a first attempt to extract the dipolar tensor components $A^{ii}$ ($K_i = A^{ii}\chi_i$, for **B**||**i**, where i denotes the crystallographic direction) we have compared [5] the Knight shift, $K_i(T)$, with the susceptibility, $\chi_i(T)$ in the standard Clogston-Jaccarino plot, $K(\chi)$. However, this plot reveals

several unusual features: (i) $K_a(\chi_a)$ deviates strongly from the expected linear behaviour, while $K_c(\chi_c)$ is approximately linear, and (ii) the direct contact contribution to the Knight shift $K_{con}^i$ ($T\to\infty$; i denotes the crystallographic direction) is strongly anisotropic, while the Pauli susceptibility $\chi_0$ ($T\to\infty$) is not. This strongly suggests that the local and bulk susceptibilities differ, which hampers the determination of the components of the dipolar tensor $A^{ii}$ in the standard fashion. We therefore followed a more straightforward method to determine the muon localization site.

It turns out that $K_a(T)$ and $K_c(T)$ follow a modified Curie-Weiss behaviour accurately, albeit in a rather limited temperature range. The lower temperature limit is about 30 K, which is also the temperature where below $\chi_a(T)$ and $\chi_c(T)$ deviate from the Curie-Weiss law, because of the stabilisation of antiferromagnetic interactions, while the upper limit (~ 100 K) is determined by muon diffusion. Therefore we write:

$$K_i(T) = K_{con}^i + A_{tot}^i C / (T - \theta_i) \qquad (2)$$

with $C=N_A\mu_0\mu_{eff}^2/3k_B$ is the Curie-Weiss constant, $\theta_i$ is the paramagnetic Curie temperature and the coupling constant $A_{tot}^i = A_{con} + A_{dip}^i$. In Fig.3 we show $K_i(T)$ where the solid lines present fits to eq.2. In this way we can determine $K_{con}^i$, $A_{tot}^i C$ and $\theta_i$. The results for $K_{con}^i$ and $A_{tot}^i C$ are listed in Table I, where we have taken $\theta_i$ equal to 100 K and 66 K, for the a and c-axis, respectively, as determined from a similar analysis on UPt$_3$ [6]. Next we have evaluated $A_{dip}^i$ from the values of $A_{tot}^i$ and $A_{con}$ (by making use of $\vec{A}_{con}=1/3\,\mathrm{Tr}(\vec{A}_{tot})\vec{E}$, where $\vec{E}$ is the unit tensor) for the electronic configurations $f^1$ (U$^{5+}$), $f^2$ (U$^{4+}$) and $f^3$ (U$^{3+}$) with effective moments $\mu_{eff}$= 2.54, 3.58 and 3.62 $\mu_B$/U-atom, respectively. The results for **B**∥**a** (frequency components $\nu_1$ and $\nu_2$) and **B**∥**c** are listed in Table II. Finally we compare $A_{dip}^i$ with the calculated values for axial symmetric sites, listed in Table III. In the case of the $f^2$ and $f^3$ configurations we find a very good agreement if the stopping site is (0,0,$z$). Symmetry considerations lead us to propose that the most likely value of z is 0, and the stopping site is (0,0,0). The location of this site in the unit cell of UPt$_3$ is shown in Fig.4. It should be noted that the $f^2$ and $f^3$ free-ion effective moment values differ from the value $\mu_{eff}$~ 2.8 $\mu_B$/U-atom determined from the modified Curie-Weiss fit to $\chi(T)$. This indicates once more that the local susceptibility differs from the bulk susceptibility. A puzzling aspect is the splitting of the Knight shift component for **B**∥**a** below ~100 K. Since our analysis shows that there is only one stopping site, the splitting has to be attributed to two spatially distinct regions of different magnetic response [7]. The origin for this lies most likely in the direct contact contribution due to the conduction electrons, given the distinctly different values of $K_{\text{con}}^i$ (see Table I) for the two frequency components $\nu_1$ and $\nu_2$.

**Acknowledgements:**

This work was part of the research programme of the Dutch "Stichting FOM" (Foundation for Fundamental Research of Matter). P.E. acknowledges the European Commission for financial support within the TMR programme. M.J.G. acknowledges support through NATO collaborative research grant CGR 11096.

**Tables and table captions**

Table I  Parameters deduced from a fit of the Knight shift of U(Pt$_{0.95}$Pd$_{0.05}$)$_3$ to the modified Curie-Weis law (see eq.2). For **B**∥ **a** the parameters for the two different signals are labelled by $\nu_1$ and $\nu_2$.

| $i$ | $K_{\text{con}}^i$ (ppm) | $A_{\text{tot}}^i C$ (K$^{-1}$) | $\theta_i$ (K) |
|---|---|---|---|
| **B**∥ **a** ($\nu_1$) | -21(30) | -0.074(4) | -66 |
| **B**∥ **a** ($\nu_2$) | -537(128) | -0.100(15) | -66 |
| **B**∥ **c** | 568(43) | -0.315(20) | -100 |

Table II  Dipolar tensor components $A_{\text{dip}}^i$ of U(Pt$_{0.95}$Pd$_{0.05}$)$_3$ for the electronic configurations $f^1$ (U$^{5+}$), $f^2$ (U$^{4+}$) and $f^3$ (U$^{3+}$).

| U | $A_{\text{dip}}^{a,\nu_1}$ (mol/cm$^3$) | $A_{\text{dip}}^{c,\nu_1}$ (mol/cm$^3$) | $A_{\text{dip}}^{a,\nu_2}$ (mol/cm$^3$) | $A_{\text{dip}}^{c,\nu_2}$ (mol/cm$^3$) |
|---|---|---|---|---|
| 3+ | 0.0039 | -0.0078 | 0.0034 | -0.0070 |
| 4+ | 0.0040 | -0.0080 | 0.0034 | -0.0071 |
| 5+ | 0.0079 | -0.0158 | 0.0068 | -0.0141 |

Table III  Calculated dipolar tensor components for several axial symmetric sites ($A_{\text{dip}}^a = A_{\text{dip}}^{a*}$). The first column gives the multiplicity and the Wyckoff letter of the particular site.

| Wyckoff letter | interstitial site | $A_{\text{dip}}^a$ (mol/cm$^3$) | $A_{\text{dip}}^c$ (mol/cm$^3$) |
|---|---|---|---|
| 2a | 0 0 0 | 0.0037 | -0.0075 |
| 4e | 0 0 1/8 | 0.0039 | -0.0078 |
| 2b | 0 0 1/4 | 0.0041 | -0.0082 |
| 4f | 2/3 1/3 0 | -0.0717 | 0.144 |
| 4f | 2/3 1/3 1/8 | -0.0215 | 0.0430 |
| 2d | 2/3 1/3 1/4 | -0.0128 | 0.0256 |

**Figure captions**

Fig.1 Temperature dependence of the Knight shift of U(Pt$_{0.95}$Pd$_{0.05}$)$_3$ (sample #2): (●) one component fit for **B**‖**a**, (■, ◆) two component fit ($\nu_1$, $\nu_2$) for **B**‖**a**, and (○) **B**‖**c**. The solid lines are to guide the eye.

Fig.2 Angular variation of the Knight shift frequency of U(Pt$_{0.95}$Pd$_{0.05}$)$_3$ (sample #1) for a field (*B*= 0.6 T) oriented in the a-c (○) and a-a* (Δ) plane measured at *T*= 10 K. The solid line represents a fit to a cos$^2$-law.

Fig.3 Knight shift of U(Pt$_{0.95}$Pd$_{0.05}$)$_3$ (sample #2) between 30 and 100 K: (■, ◆) frequency components ($\nu_1$, $\nu_2$) for **B**‖ **a**, and (○) **B**‖ **c**. The solid lines represent fits to eq.(2).

Fig.4 The crystallographic unit cell of UPt$_3$ (space group P6$_3$/mmc). The axial symmetric sites (0,0,0), (2/3,1/3,0) and (2/3,1/3,1/4) are indicated.

Figure 1

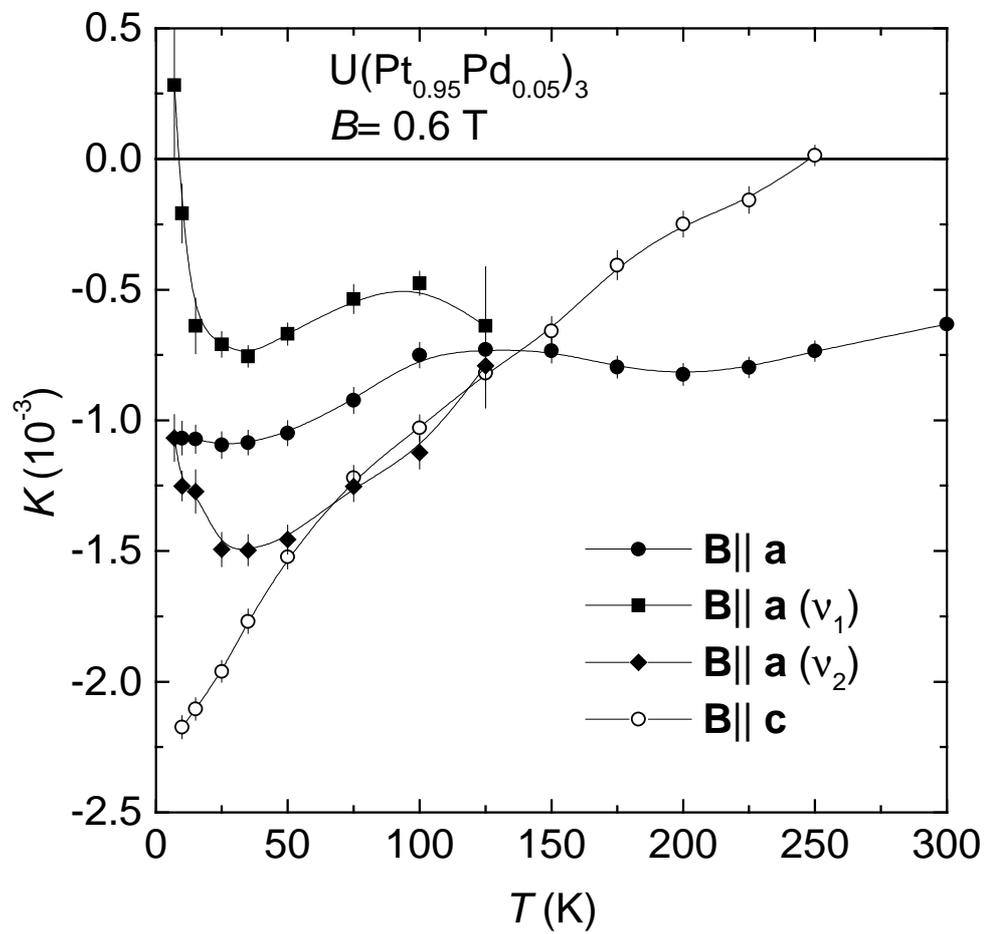

Figure 2

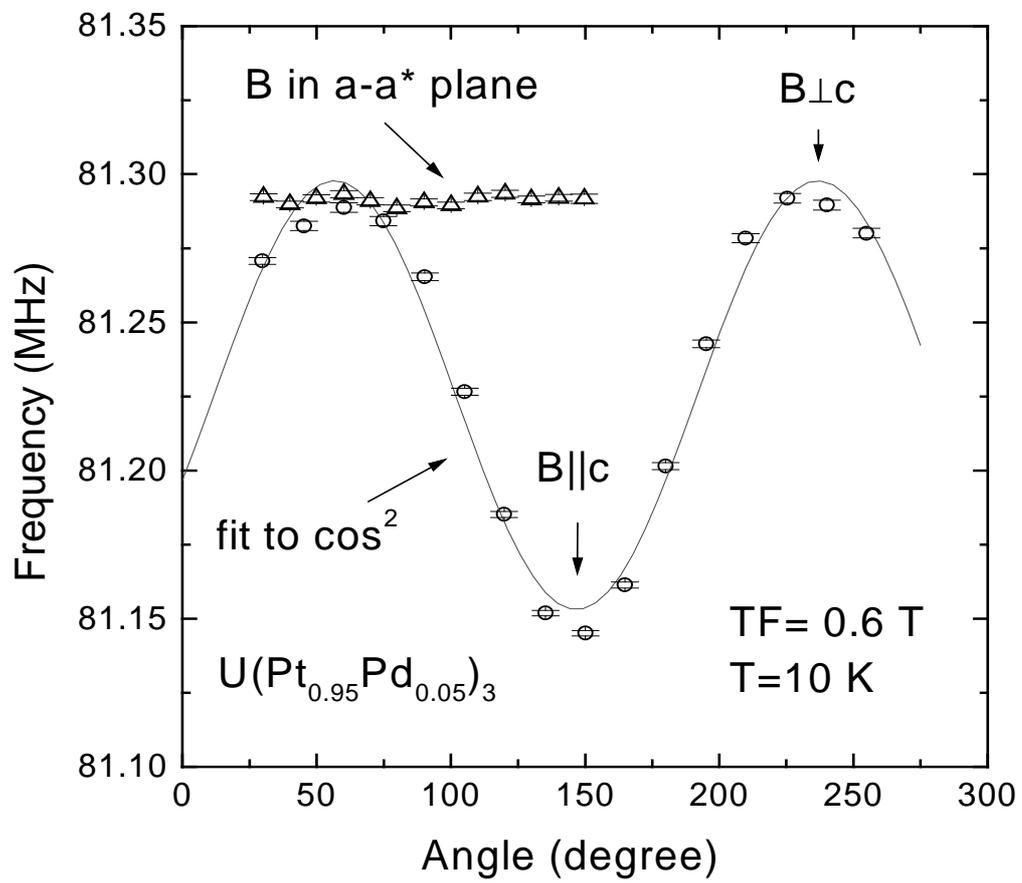

Figure 3

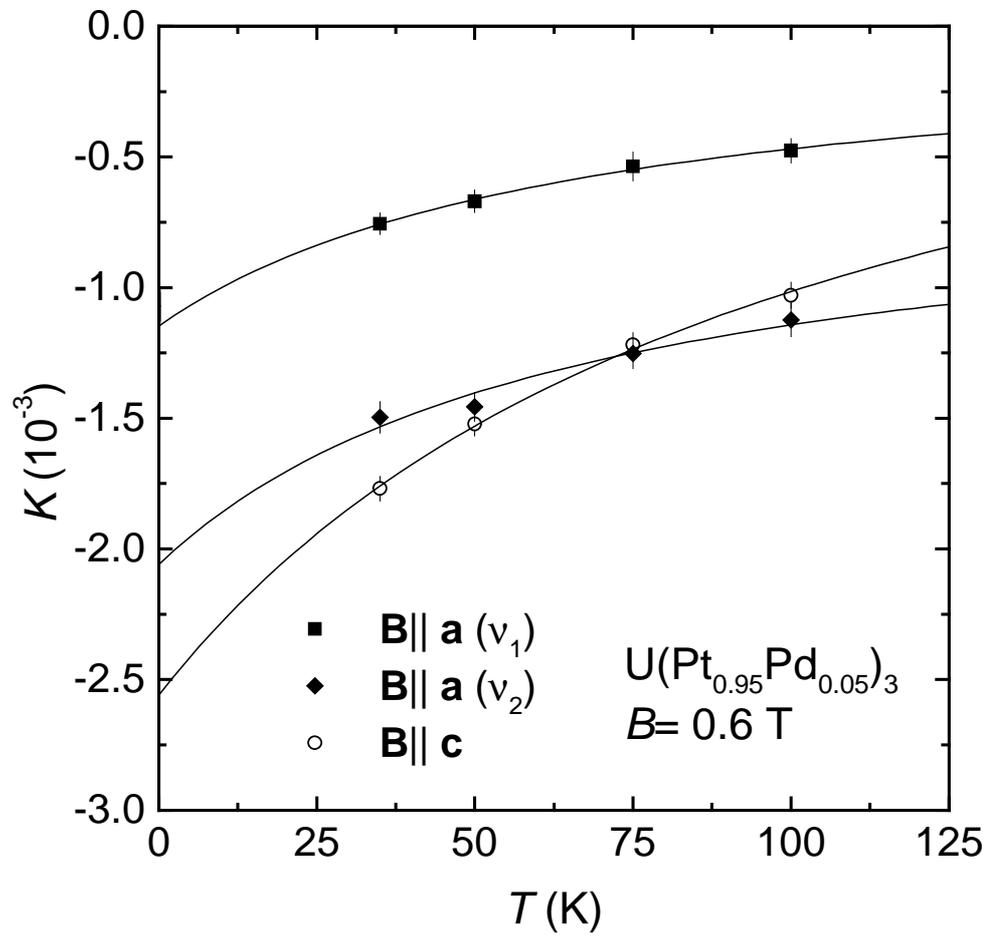

Figure 4

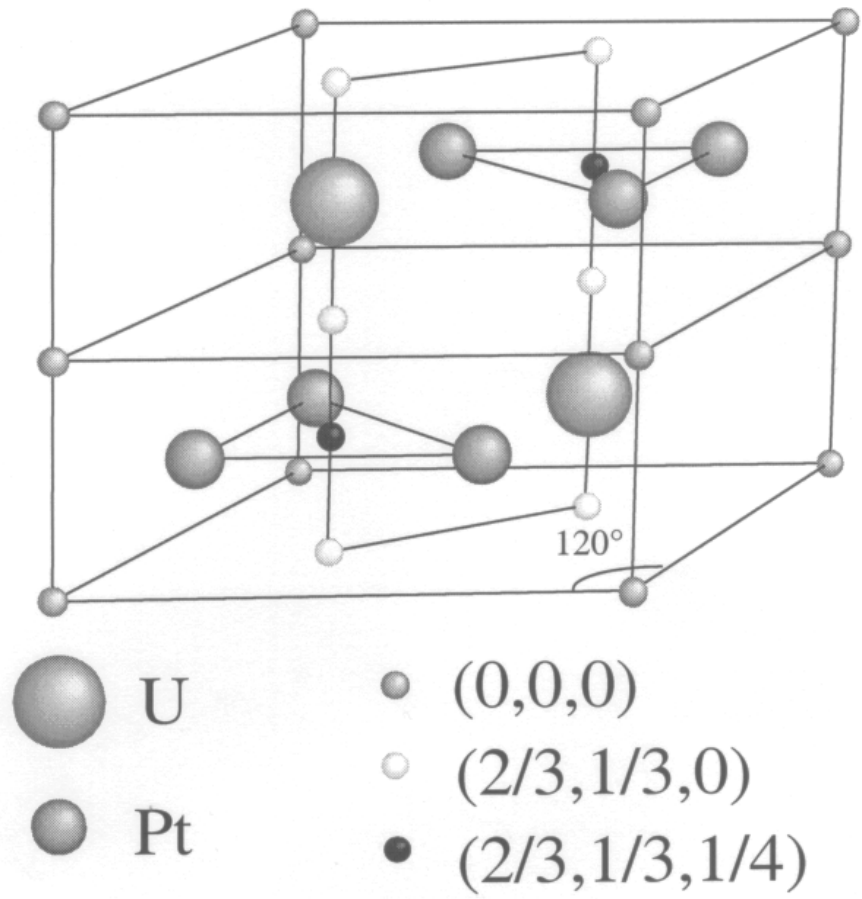

- U
- Pt
- (0,0,0)
- (2/3,1/3,0)
- (2/3,1/3,1/4)